# Spin injection across a III-V/chiral perovskite interface enabling spin accumulation at room temperature

Matthew P. Hautzinger[1], Xin Pan[2], Steven C. Hayden[1], Jiselle Y. Ye[1,3], Qi Jiang[1], Mickey J. Wilson[1], Yifan Dong[1], Emily K. Raulerson[1], Ian A. Leahy[1], Chun-Sheng Jiang[1], Joseph M. Luther[1], Yuan Lu[4], Katherine Jungjohann[1], Z. Valy Vardeny[2], Joseph J. Berry[1,5,6], Kirstin Alberi[1], Matthew C. Beard[1]

[1]National Renewable Energy Laboratory, Golden, Colorado 80401, United States

[2]Department of Physics & Astronomy, University of Utah, Salt Lake City, UT 84112, USA

[3]Department of Physics, Materials Science Program, Colorado School of Mines, Golden, Colorado 80401, USA

[4]Institut Jean Lamour, Université de Lorraine, CNRS, UMR 7198, 54011 Nancy, France

[5]Renewable and Sustainable Energy Institute, University of Colorado Boulder, Boulder, CO 80309, USA

[6]Department of Physics, University of Colorado Boulder, Boulder, CO 80309, USA

**Abstract**

Spin accumulation in semiconductor structures at room temperature and without magnetic fields is key to enable a broader range of opto-electronic functionality. Current efforts are limited due to inherent inefficiencies associated with spin injection into semiconductor structures. Here, we demonstrate spin injection across chiral halide perovskite/III-V interfaces achieving spin accumulation in a standard semiconductor III-V $(Al_xGa_{1-x})_{0.5}In_{0.5}P$ multiple quantum well (MQW) light emitting diode (LED). The spin accumulation in the MQW is detected via emission of circularly polarized light with a degree of polarization of up to ~15%. The chiral perovskite/III-V interface was characterized with X-ray photoemission spectroscopy (XPS), cross sectional scanning Kelvin probe force microscopy, and cross section transmission electron microscopy (TEM) imaging, showing a clean semiconductor/semiconductor interface where the fermi-level can equilibrate. These findings demonstrate chiral perovskite semiconductors can transform well-developed semiconductor platforms to ones that can also control spin.

Semiconductors are central to modern society because they allow for manipulation of charge, but controlling spins in semiconductor structures at room temperature remains elusive. Nevertheless, control over the spin-degrees of freedom is the basis for a multitude of commercial products central to current day society, most notably for magnetic memory,[1] but is achieved through forming heterostructures between ferromagnets (FM) and normal metals, resulting in, e.g., the giant magnetoresistance effect[2,3] and the tunnelling magnetoresistance effect.[4] However, despite the potential applications such an approach using semiconductors, i.e., semiconductor/FM interfaces has been limited.[5] Chiral induced spin selectivity (CISS) describes the spin dependent transmission of charge carriers through an oriented chiral potential, where the resulting spin orientation is parallel to the chiral helicity, i.e., the chiral structure determines the spin-orientation.[6–8] CISS has most commonly been observed in self-assembled chiral molecular layers and thus is limited for integration with common semiconductors. Recently, a family of chiral-semiconductors based upon halide perovskite semiconductors (c-HP) was discovered to exhibit CISS [9–12] and shown to enable spin accumulation in an adjacent halide perovskite light emitting layer.[13] These results demonstrate that CISS allows for spin control at room temperature, providing a pathway towards transforming modern optoelectronics, if the c-HP can be successfully incorporated. Here, we demonstrate this potential by coupling c-HP semiconductors with a standard III-V light emitting diode optoelectronic structure.

Halide perovskites are an emerging class of semiconductors being explored for optoelectronic applications that interconvert charge and light, e.g. solar cells and LEDs.[14,15] Chirality is introduced into this family of semiconductors through the incorporation of chiral organo-ammonium cations, e.g., R/S-MBA (MBA = $\alpha$-methylbenzylammonium). The resulting chiral semiconductors consists of highly ordered bilayers of the chiral ammonium cation interceding corner sharing metal-halide octahedra sheets.[16,17] The chirality of the organic component imbues chiral behavior to the hybrid system. This family of chiral semiconductors have two important characteristics that make them unique candidates for controlling spin in semiconductor platforms: (1) their enantiomeric purity is deterministically selected by the choice of organic constituent chirality (R or S) and (2) upon spin-coating they organize themselves into highly textured, crystalline thin films where the chiral organic molecules are perpendicular to the substrate akin to stacked self-assembled monolayers. Therefore, c-HP can in principle be incorporated into semiconductor structures, heterostructures, and optoelectronics.[13,18,19] In this work, we show that c-HP can in fact be easily integrated with a III-V LED structure to transform it to a spin-LED. Spin-polarized carriers are injected into the conventional III-V LED structure and the resulting spin-accumulation in the III-V is detected via the interconversion of spin-polarized carriers into circular polarized light, through the conservation of angular momentum. The resulting circularly polarized light emission is a five-fold increase over previous CISS injection,[13] a result of the successful integration c-HP with III-V structures.

We fabricated c-HP/SC spin injection interfaces designed to inject spin polarized holes into commercially relevant $(Al_{0.32}Ga_{0.68})_{0.5}In_{0.5}P$ multiple quantum well (MQW) LED structures (Fig. 1a).[20] The AlGaInP base LED structure was received as a completed LED structure without electrical contacts and the internal structure was not modified. The structure consist of an n-doped GaAs substrate (Si-doped ~$10^{18}$ cm$^{-3}$) with a layer stack (from bottom up) arranged as follows: 200 nm n-type $(Al_{0.53}Ga_{0.47})_{0.5}In_{0.5}P$ cladding layer (Si-doped $1\times10^{18}$ cm$^{-3}$); 5x $(Al_{0.32}Ga_{0.68})_{0.5}In_{0.5}P$ multiple quantum wells (QW, 2.10 eV bandgap, nominally undoped, 10 nm thick) within $(Al_{0.48}Ga_{0.52})_{0.5}In_{0.5}P$ barriers (QB, 2.20 eV bandgap, nominally undoped, 20 nm thick); 200 nm p-type $(Al_{0.53}Ga_{0.47})_{0.5}In_{0.5}P$ cladding layer (Zn-doped ~$1\times10^{17}$ cm$^{-3}$). The device structure was capped with a ~50 nm GaAs layer for protection, which was removed by chemical etching (2:1:10 $NH_4OH:H_2O_2:H_2O$) immediately prior to adding the c-HP layers. $(R/S-MBA)_2PbI_4$ as the spin injector (thickness ~110 nm $\pm$ 6nm) and TFB (hole transport material) were spin coated on top of the pristine $(Al_{0.53}Ga_{0.47})_{0.5}In_{0.5}P$ cladding layer surface. To improve the wetting of the AlInGaP surface and ensure conformal contact we found it necessary that $(R/S-MBA)_2PbI_4$ was spin coated twice. Transparent conducting layers (aluminum (III) oxide and indium doped zinc (II) oxide (IZO)) and gold top and bottom contacts were then deposited. Detailed description is provided in the supplementary information. Top-down and cross-sectional SEM of each deposited layer is shown in Fig. S1.

The spin accumulation observed here was enabled by the direct formation of a c-HP semiconductor interface with III-V semiconductor (Fig. 1). The c-HP deposited directly on the III-V produces an abrupt SC/SC interface. Cross sectional TEM (Fig. 1b and c) samples were prepared by room temperature milling and lift out of the AlInGaP/C-HP heterostructure. The sample was then kept under cryogenic conditions during transfer and imaging (details in the SI). These images show the SC/SC interface between the (R/S-MBA)$_2$PbI$_4$ and AlGaInP LED. The c-HP, when spin-coated, conformally contacts the III-V semiconductor with no noticeable voids or visible imperfections. These are the first images to our knowledge of a HOIS/III-V interface without oxide barriers present. Spin coating twice (discussed above) was essential for producing high quality pin hole free films and prevents delamination during the cross-sectional TEM imaging.

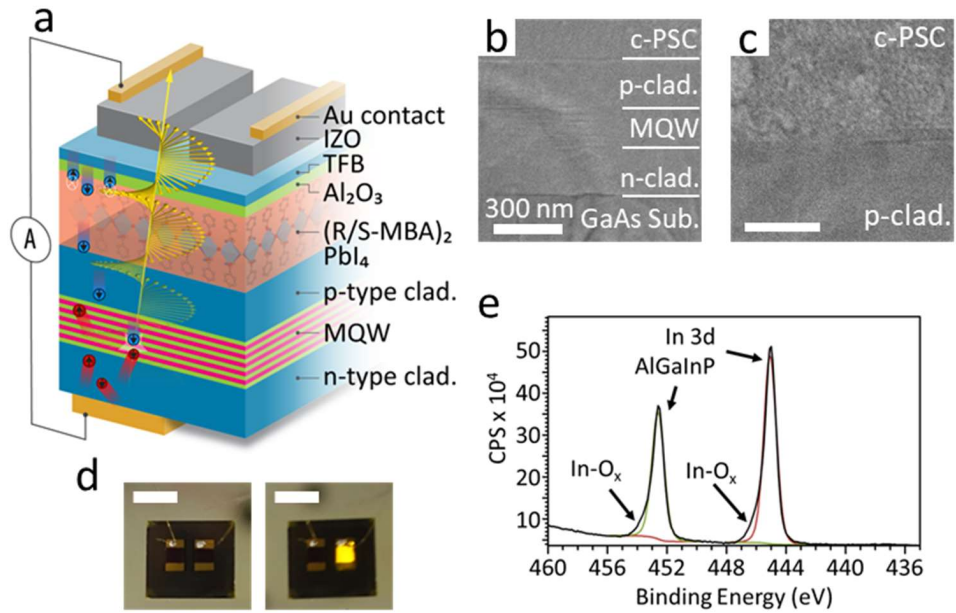

**Fig. 1. Device schematic and interface characterization.** (a) Device stack of the spin LED emitting circularly polarized electroluminescence (CP-EL). The (R-MBA)$_2$PbI$_4$ acts as a spin filter, allowing only spin polarized holes (blue circles) to flow through the device and recombine in the MQWs emitting CP-EL (yellow helix). (b-c) Cross sectional TEM of the AlGaInP/(S-MBA)$_2$PbI$_4$ interface. No structures in between the perovskite and AlGaInP can be observed. (d) Images of the device off and on showing the yellow electro luminesce. Scale bars are 0.5 cm. (e) XPS of the etched AlGaInP surface.

XPS of the AlGaInP (Fig. 1e) prior to c-HP deposition shows only slight evidence of an oxide forming on the AlGaInP substrate indicated by the low broadening of the In peaks. The broadening that does exist may be the result of native oxide formation during transport of the sample from an ambient environment to the XPS. This low oxide content is essential for producing efficient spin injection across the interface. Further XPS characterization (Fig. S2) suggests the surface is InP and GaP terminated, which may be helpful for the c-HP wetting during spin coating. This oxide-free surface is enabled by the removal of the protective GaAs capping layer from the AlGaInP substrate LED by chemical etching immediately prior to (R/S-MBA)$_2$PbI$_4$ deposition. Thus, due to the conformal contact and oxide free interface we expect that carriers can equilibrate across this interface with minimal loss, i.e., the c-HP layer behaves as any other SC layer within a device stack and the properties are controlled by the SC/SC interface, not oxide barriers.

The resulting LED (Fig. 1d) shows light emission across the active area (~ 4 mm$^2$). The light emission intensity increases linearly with applied current (Fig. S3 a, b). I-V curves of the devices show diode-like behavior, with a slightly higher onset voltage than without c-HP incorporation, due to (R/S-MBA)$_2$PbI$_4$/AlGaInP band offset (Fig. S3 and discussed below). To demonstrate controlled spin-accumulation in the III-V LED structure we measured the resulting circularly polarized electroluminescence (CP-EL). The CP-EL was measured through a quarter waveplate and linear filter to detect left-handed vs. right-handed circular polarization intensity. Average CP-EL spectra shown in Fig. 2a and 2b show the prototypical MQW device emission with center wavelength at 590 nm. The emission intensity is greater for right-handed emission when (S-MBA)$_2$PbI$_4$ is employed while the intensity is greater for left-handed emission when (R-MBA)$_2$PbI$_4$ is employed. The degree of circular polarization (DOCP) is defined as the difference between the intensity of right handed ($I_{right}$) and left-handed ($I_{left}$) circularly polarized light over the sum of both intensities (equation 1).[13]

Equation 1. $P_I = (I_{left} - I_{right})/(I_{right} + I_{left})$

The DOCP emission reaches > 10% when (S-MBA)$_2$PbI$_4$ and is <-10% when (R-MBA)$_2$PbI$_4$ is incorporated into the device. The polarized emission is a direct consequence of spin-accumulation in the III-V MQW structure during the device operation and is correlated to the handedness of the c-HP, no light is emitted from the c-HP, it only serves to deliver spin-polarized carriers to the non-chiral emitter layer.

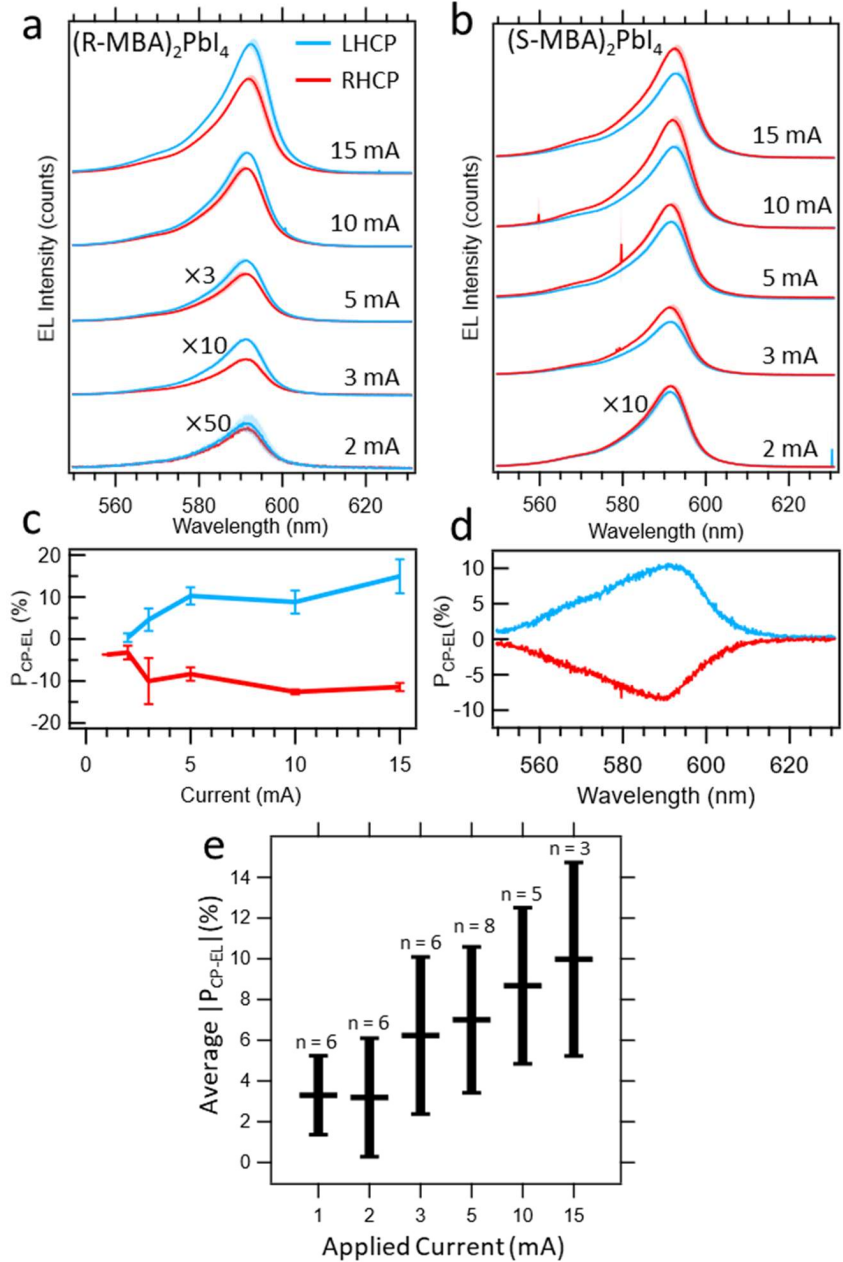

**Fig. 2. CP-EL emission of (R/S-MBA)$_2$PbI$_4$/(Al$_x$Ga$_{1-x}$)$_{0.5}$In$_{0.5}$P devices.** Emission of left and right-handed light from (a) (R-MBA)$_2$PbI$_4$ and (b) (S-MBA)$_2$PbI$_4$. (c) Degree of circular polarization (DOCP) of these devices as a function of current applied. (d) P$_{CP-EL}$ of the devices as a function of wavelength (at 5 mA). (e) Average polarization

of several devices with applied current. Typical voltages at 1 mA and 15 mA are 5.5 V and 7.2 V, respectively. The n-value corresponds to the number of devices averaged with the specific applied current.

The DOCP at each drive current are shown in Fig. 2c. The DOCP achieved is ~10% across multiple LEDs when the drive current exceeds 5 mA and reaches up to 15% for the best device at the highest drive currents. The DOCP remains strong across the 565-600 nm range (Fig. 2d). For statistical purposes, additional sample LEDs are shown in the supplementary information. In addition, we performed a *pseudo in situ* experiment, where the LED is continuously operated and the quarter waveplate was rotated selecting for RH and LH circular polarization (Fig. S4) over multiple cycles with continuous collection. While the overall EL intensity decreases with time due to joule-heating, the difference in intensity between RH and LH CP-EL remains > 10%.

A DOCP >15% implies a much higher spin accumulation since DOCP is a product of the spin accumulation multiplied by the circular polarization efficiency in the III-V MQWs, which is estimated to have a maximum of ~ 50%.[21] Therefore, the spin transfer process is highly efficient. To further confirm that the CP-EL is a result of spin polarized accumulation into the SC we measured the DOCP under an applied magnetic field (i.e. optical Hanle effect). Application of external magnetic field leads to spin presession around the field direction, causing a decrease in the DOCP (Fig. S5).[22,23] In our measurement, a device with 6.5% DOCP at 0 mT decreases to ~4% at 130 mT. The decrease in DOCP under applied magnetic fields confirms that the CP-EL is the result of spin accumulation within the emitter MQW layers

The average DOCP (Fig. 2e) is found to increase with increasing drive currents (and corresponding voltage) and this characteristic is consistent across multiple devices. Our observations are similar to DOCP bias dependence observed in traditional III-V spin-LEDs where the DOCP increases at higher bias (current), up to a maximum, then decreased with further bias, but is opposite to the behavior observed in organic spin-LEDs where the DOCP decreases with higher applied bias.[24] In the III-V reports, the bias dependence was suggested to arise from the interplay between positively impacting electric fields resulting in increasing spin lifetimes in the emitter layers at moderate biases but with enhanced D'yakonov Perel' (DP) spin relaxation reduces the DOCP at higher biases.[25–27] Testing of our devices at higher drive currents was not possible, but the spin-physics within the III-V should be similar.

We believe the DOCP is initially low at low bias due to the barriers from poorly aligned type-II bands (Fig 3a) at the c-HP/AlGaInP interface. When the carriers equilibrate (at 0 V bias, Fig. 3b), there is a large depletion region into the c-HP with low carrier concentration, which allows for the formation of a 2D hole gas within the c-HP at the interface. A large depletion region with intermittent carriers provides opportunities for spin scattering during spin-injection.[28] However, a higher bias (3V is shown in Fig. 3b) collapses the depletion region and gives injected carriers more energy that reduces interactions at the interface. Similar mechanisms of spin scattering due to depletion region and interface barrier heights have been suggested previously.[27,29]

To confirm the conjectured band bending and carrier equilibration, we performed cross-sectional kelvin force probe microscopy (KPFM) (Fig. 3c). KPFM measures the surface potential of a cleaved samples cross section by probing the coulomb force between the probe and sample.[30] The surface potential of the cross-sections measured by KPFM is dominated by electrical charges trapped at the exposed cross section surface states, as such a bias voltage is applied across the device between the electrical contacts. Varying the bias ensures (0.5 to -1.5 V) ensures removal of the traps, while ensuring surface charges don't move under the applied bias. In the electric field subplot (derivative of potential difference with respect to distance) of Fig 3c., we observe the presence of a larger depletion region in the c-HP and smaller region in the p-clad AlGaInP adjacent to the c-HP in agreement with our calculated band alignments. The cross-sectional KPFM further demonstrates that we have formed a SC-SC interface with band bending, confirming conformal contact and indicating our interface is not impeded by extra structures (e.g. oxides). Tuning the c-HP energy levels relative to the SC could eliminate this barrier and lead to greatly improve injection efficiency.

There is considerable effort in coupling HP with traditional semiconductor platforms (Si, GaAs, CdTe) to produce tandem solar cells. Despite this interest there are very few reports of HP semiconductors (including c-HPs) being integrated with a direct contact on top of conventional SCs in electronic devices. Most approaches rather employ a contact and/or transport layers between the HP and SC. Here we have demonstrated direct contact of the HP with traditional semiconductors is possible and that the HP semiconductor behaves as another semiconductor within the device stack. We believe the challenge for the forming direct contact is in maintaining an oxide free conventional SC surface prior to direct spin-coating the HP.

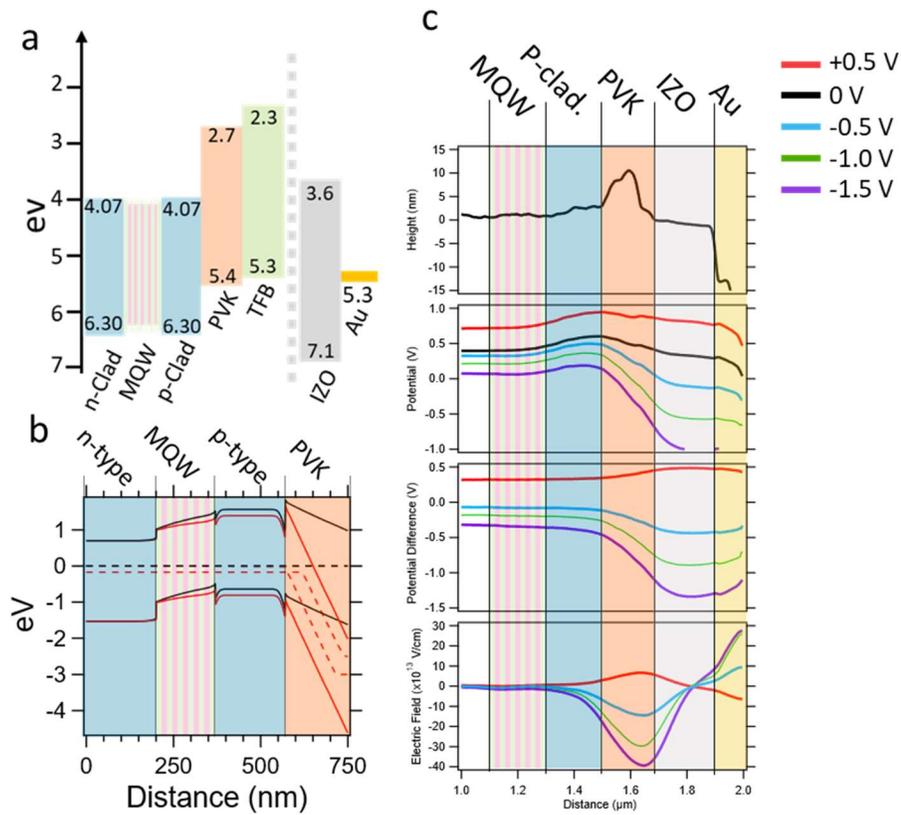

**Fig. 3. Further characterization of the CP-EL** (a) band alignments determined from literature UPS values. (b) Band bending at 0V and 3V determined by semiconductor device simulations. Dashed lines show the electron/hole quasi-fermi level splitting. (c) Cross-sectional Kelvin probe force microscopy and AFM of the spin LED. The electric field drop is seen across the (R-MBA)$_2$PbI$_4$ (labeled PVK) in good agreement with our calculated energy band diagrams.

A high spin-accumulation in conventional SCs at room temperature and no applied magnetic field has been difficult to achieve due to the required SC/spin injector interface and subsequent spin scattering at the interface.[31,32] Directly contacting the semiconductor with a ferromagnetic (i.e., SC/FM interface) where the spin polarization can be controlled through the magnetic polarization of the FM suffer from a conductivity mismatch, $\gamma \ll 1$. A tunnel oxide barrier is required to achieve efficient spin injection. These interfacial architectures are hindered by undesired traits such as intrinsic-defects that decrease the spin injection efficiency.[25,29,33–35] For the c-HP/SC developed here the conductivity mismatch is no longer an issue as the conductivity of the c-HP[36] is lower than the p-clad SC layer receiving the injection (opposite of a FM/SC interface) and similar to diluted magnetic semiconductor (DMS) spin injection systems.[31] In fact, the best spin injection analog to c-HP are DMS, which have the advantage of forming a SC/SC interface with no conductivity mismatch issue, allowing efficient spin injection. However, typical DMS have curie temperatures well below RT, which inhibits their utility. Most systems that exhibit CISS, on the other hand, are also not suitable and have not been used for spin injection into conventional SCs (e.g. III-Vs, Si) because most CISS systems are insulating and consists of molecular layers. The c-HP system seems to incorporate the best situation of high spin injection at room temperature and compatibility with existing semiconductor platforms.

In this study, the Spin-LED platform is used to detect spin polarized current without the need for ferromagnetic contacts, as is commonly done in the CISS literature. This could serve as a platform for studying CISS in the absence of FMs. The spin-LED is also an interesting technological device with potential for advanced applications, such as circularly polarized single photon sources information processing, and biological imaging.[37–39] The CISS based approach is quite efficient and attractive for spin-LED applications due to several reasons. First, the c-HP is known to produce a highly spin polarized current, forming a basis for efficient spin injection. The spin orientation produced via the CISS mechanism is parallel to the direction of current and the light emission is also parallel to the current direction due to the small escape cone of the III-V. This promotes spin-to-light conversion and operate as a true vertical stack instead of require edge emission because optical selection rules require the circularly polarized light helicity be parallel to the emission direction.[33,40] In other spin injection contacts the produce spin polarized currents in specific orientations that may not be desirable for the directionality of light emission or inconvenient for charge manipulation.[33] For spin-LEDs in particular, spin injection contacts must be designed to not block or reabsorb electroluminescence.

Integration of c-HOIS transforms an existing commercially relevant III-V LEDs from a conventional LED semiconductor structure that controls the interconversion of light and charge to one that now also controls light-to-spin. Thus, our approach produces a functional spinbased semiconductor structure operating at room temperature with no external magnetic fields. The high spin injection efficiency is a result of the c-HP/III-V interface we have developed, where TEM, XPS, and KPFM indicate a direct SC-SC interface, which allows carrier equilibration and efficient spin injection. This demonstration shows that FMs and other previous spin injectors that require an external magnetic field are not required for spin functionality. The c-HOIS SC/SC interface developed here forms the basis of a new class of spin injectors to achieve spin-accumulation for a variety of spin functionalities. In addition, it provides a well-behaved platform for investigating CISS. Until now, the limited devices integrating chiral molecules has hindered research investigations into basic CISS behavior, and CISS/SC heterostructures will be a valuable tool.

**Funding:**

This work was supported as part of the Center for Hybrid Organic Inorganic Semiconductors for Energy (CHOISE) an Energy Frontier Research Center funded by the Office of Basic Energy Sciences, Office of Science within the U.S. Department of Energy (DOE). This work was authored in part by the National Renewable Energy Laboratory, operated by Alliance for Sustainable Energy, LLC, for DOE under Contract No. DE-AC36-08GO28308. The views expressed in the article do not necessarily represent the views of the DOE or the U.S. Government. Support for structural and microscopy characterization and device characterization was provided by a Laboratory Directed Research and Development project funded by the National Renewable Energy Laboratory.

**Author Contributions:** M.P.H. M.C.B., K.A, J.M.L., J.B., Y.L., conceived of the research idea and designed experiments. M.P.H. fabricated the devices and measured the CP-EL. M.W. deposited the IZO. J.Y. performed XPS. Q.J. and I.A.L. aided in the device fabrication process and basic device characterizations. Y.D. and E.K.R. performed spectroscopic characterization. C.S.J. performed cross-sectional KPFM. X.P. and V.Z.V. performed and interpreted Hanle effect measurements.

**Supplementary materials**

Materials and Methods

Figs S1 to S7

References (1-6)